\documentclass[aps,prl,a4paper,showpacs,twocolumn,floatfix]{revtex4}
\usepackage[latin1]{inputenc}
\usepackage{amsfonts}
\usepackage{amssymb}
\usepackage{amsmath}
\usepackage{calc}
\usepackage{graphicx}
\usepackage{epsfig}
\usepackage{bm}
\usepackage{ulem}
\usepackage{setspace}
\usepackage{subfigure}

\begin{document}

\title{Long-Distance Entanglement Distribution with Single-Photon Sources}

\date{\today}
\author{Nicolas Sangouard}
\author{Christoph Simon}
\author{Ji\v{r}\'i Min\'a\v{r}}
\author{Hugo Zbinden}
\author{Hugues de Riedmatten}
\author{Nicolas Gisin}
\affiliation{Group of Applied Physics, University of
Geneva, Switzerland}

\begin{abstract}
We present an efficient architecture for quantum repeaters
based on single-photon sources in combination with quantum
memories for photons. Errors inherent to previous repeater
protocols using photon-pair sources are eliminated, leading
to a significant gain in efficiency. We establish the
requirements on the single-photon sources and on the photon
detectors.
\end{abstract}

\maketitle

Entangled state distribution over long distances is a
challenging task due to the limited transmission efficiencies
of optical fibers. To overcome this problem, quantum
repeaters are likely to be required \cite{Briegel98}. The
basic principle of quantum repeaters consists in
decomposing the full distance into shorter elementary
links. Quantum memories allow the creation of entanglement
independently for each link. This entanglement can then be
extended to the full distance using entanglement swapping.

The protocol proposed here is similar to the well-known
Duan-Lukin-Cirac-Zoller (DLCZ) scheme \cite{Duan01}, and to
its recent modification based on photon pairs and
multi-mode memories (P$^2$M$^3$) \cite{Simon07}, in that
entanglement for an elementary link is created by the
detection of a single photon. However, both protocols
 rely on sources that create {\it correlated
pairs} of excitations, namely one atomic 
excitation and one photon in the case of DLCZ, and 
two photons in the case of P$^2$M$^3.$ 
These correlations allow one to
establish entanglement between distant memories based on
the detection of a photon which could have come from either
of two remote sources.

Our protocol uses {\it single-photon} sources making it possible to eliminate errors
due to two-pair emission events, which are unavoidable for
Refs. \cite{Duan01,Simon07}. This leads to a
significant improvement in the achievable entanglement
distribution rate. Moreover our scheme is compatible with
the use of multi-mode memories \cite{Simon07}, spatial and
frequency multiplexing \cite{Collins07}, and improved 
entanglement connection \cite{Jiang07}, all of which promise
additional speed-ups. \\

\begin{figure}[hr!]
{\includegraphics[scale=0.28]{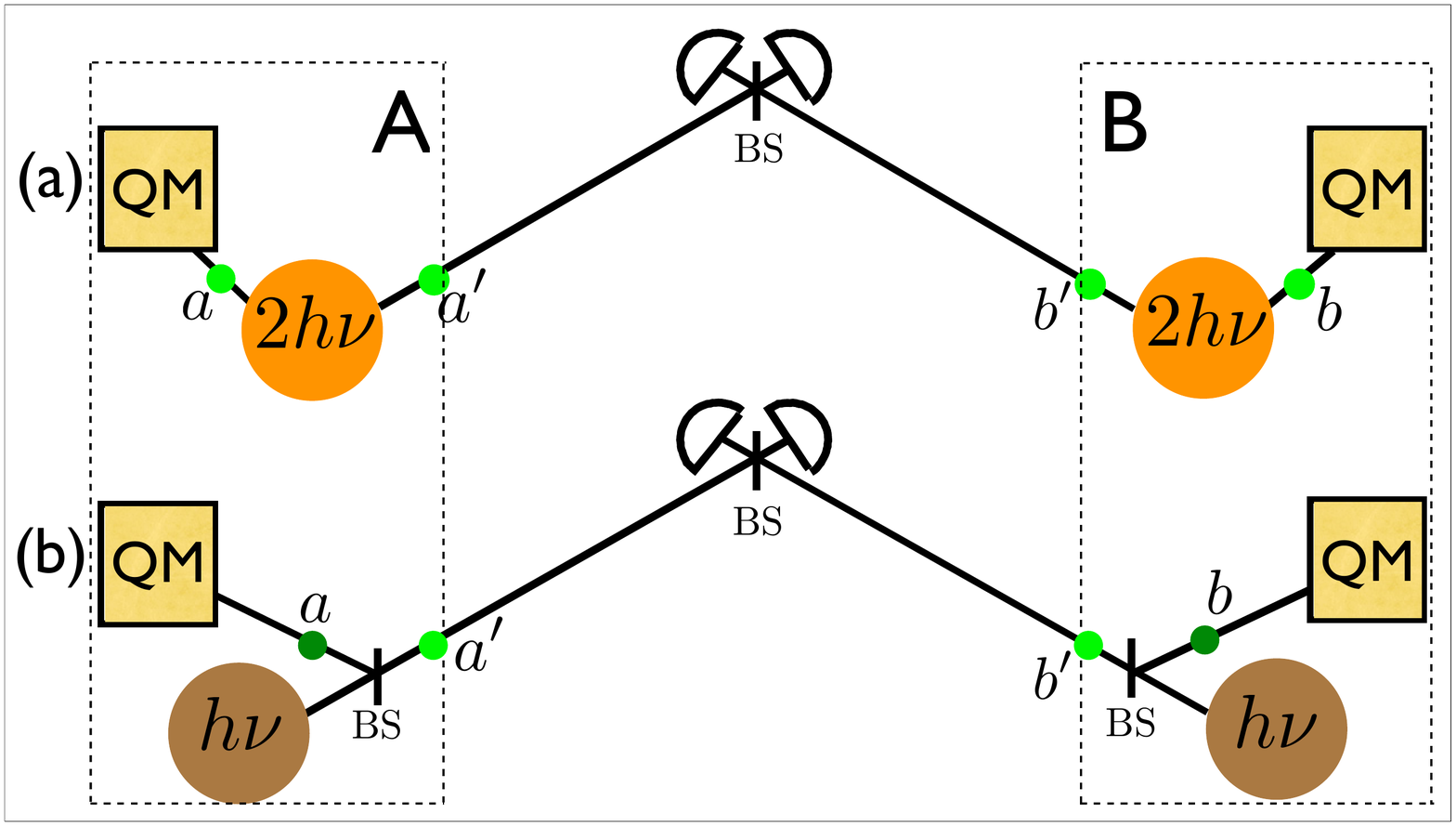}} \caption{(Color
online) Schematic architecture of an elementary link
connecting two locations $A$ and $B$ for (a) the
P$^2$M$^3$ protocol that uses photon pair sources
\cite{Simon07}, and (b) the new single-photon source
protocol. Sources, memories and detectors are represented by circles, squares and half-circles respectively. Vertical bars labeled BS denote beam-splitters. For both (a) and (b), the detection of a single photon behind the central beam-splitter projects the two memories into an entangled state.} \label{fig1}
\end{figure}

We begin by recalling the basic principles of the
P$^2$M$^3$ protocol. The DLCZ protocol is equivalent for
the purposes of the present discussion. The architecture of an 
elementary link is represented in Fig. \ref{fig1}
(a). The procedure to entangle two remote locations $A$ and
$B$ requires one photon-pair source and one memory at each
location. The pair sources are coherently excited such that
each of them can emit a pair with a small probability
$p/2,$ corresponding to the state 
\begin{equation}
\left(1+\sqrt{p/2}(a^\dag
a'^\dag+b^\dag b'^\dag)+O(p)\right)|0\rangle;
\label{p2m3}
\end{equation}
$a,$ $a'$ and $b,$
$b'$ are the pairs of modes emitted by the sources located at
$A$ and $B$ respectively, and $|0\rangle$ is the vacuum
state. The modes $a$ and $b$ are stored in memories close to the 
respective sources while the modes $a'$ and $b'$ are sent through optical
fibers to a station located half-way between $A$ and $B$,
where they are combined on a beam-splitter. Omitting for
simplicity the phase acquired by the photons during their transmission, 
the modes after the beam-splitter are
$\tilde{a}=(a'+b')/\sqrt{2}$, $\tilde{b}=(a'-b')/\sqrt{2}$ 
(see Ref. \cite{Simon07} for a more complete discussion).
The detection of a single photon in mode $\tilde{a}$,
for example, creates the state $\frac{1}{\sqrt{2}}(a^\dag+b^\dag)|0\rangle$
which corresponds to a single delocalized excitation. The
modes $a$ and $b$ are stored in memories, and the stored
single excitation can be written as an entangled state
$\frac{1}{\sqrt{2}}(|1_{\rm A}0_{\rm B}\rangle+|0_{\rm A}1_{\rm B}\rangle)$,
where $|0_{\rm A(B)}\rangle$ and $|1_{\rm A(B)}\rangle$
denote zero and one photon respectively stored in the
memory at $A$ ($B$). This entanglement can further be
extended to long distances using entanglement swapping
\cite{Duan01,Simon07}.

The performance of the described protocol, which has many
attractive features, is limited by a fundamental error
mechanism. Even if the pair sources are ideal, i.e. even if
they emit at most one pair each, there is a probability
$p^2/4$ that two pairs will be emitted in total. If this is
the case, and if one photon is lost during its transmission
through the fiber or by detector failure, the detection of
a single photon in the mode $\tilde{a}$ or $\tilde{b}$
generates the state $|1_{A}1_{B}\rangle$, corresponding to
two full memories. This state, which is not the desired
entangled state, introduces errors and limits the fidelity
of the created entanglement. To preserve high fidelity, one has to use
sources with low emission probability $p\ll1,$ such that the
probability to get simultaneous emissions at $A$ and $B$ is
sufficiently small. This limits the achievable distribution
rate of entangled states. The same problem occurs for the
DLCZ protocol, which, for the purpose of the present
discussion, differs from the P$^2$M$^3$ protocol only by
the fact that the modes $a$ and $b$ are created directly in
the memories \cite{Duan01,Simon07}.

The proposed new scheme using single-photon sources is free of these
fundamental errors. The architecture of our scheme is
represented in Fig. \ref{fig1} (b). The two remote
locations contain each one single-photon source and one
memory. When they are excited, each
of the two sources ideally creates one photon. The photons
created at $A$ and $B$ are sent through identical beam
splitters with reflection and transmission coefficients $\alpha$ and $\beta$ satisfying $|\alpha|^2+|\beta|^2=1,$ such that after the beam-splitters, the state of
the two photons is $(\alpha a^{\dag}+\beta
a'^{\dag})(\alpha b^{\dag}+\beta b'^{\dag})|0\rangle$,
which can be rewritten as
\begin{equation}
\left(\alpha^2 a^{\dag}b^{\dag}+\alpha\beta
\left(a'^{\dag}b^{\dag}
+a^{\dag}b'^{\dag}\right)+\beta^2a'^{\dag}b'^{\dag}\right)|0\rangle.
\label{2photons}
\end{equation}
The modes $a,$ $b$ are stored in memories. The modes $a'$,
$b'$ are coupled into optical fibers and combined on a beam
splitter at a central station, with the modes after the
beam-splitter denoted by $\tilde{a}$ and $\tilde{b}$ as
before. We are interested in the detection of one photon,
for example in the mode $\tilde{a}$. Let us consider the
contributions from the three terms in Eq. (\ref{2photons}).
The term $a^{\dag}b^{\dag}|0\rangle$ which corresponds to 
two full memories, cannot generate the
expected detection and thus does not contribute to the
entanglement creation. The term
$(a'^{\dag}b^{\dag}+a^{\dag}b'^{\dag})|0\rangle$ may induce the
detection of a single photon in mode $\tilde{a}$ with
probability $\alpha^2\beta^2  \eta_t \eta_d$, where
$\eta_{t}$ is the efficiency of transmission to the central station, 
and $\eta_d$ is the single-photon detection efficiency. For this term,
the detection of a photon in $\tilde{a}$ creates the
desired state $\frac{1}{\sqrt{2}}(a^{\dag}+b^{\dag})|0\rangle$ associated to entangled memories. 
Note that in contrast with the P$^2$M$^3$ protocol, the entanglement creation uses correlations 
between modes $a'$-$b$ and $a$-$b'$ rather than correlations between $a$-$a'$ and $b$-$b'$ (see relevant term in Eq. (\ref{p2m3})). Finally the
term $a'^{\dag}b'^{\dag}|0\rangle$ may also produce a
single photon in mode $\tilde{a}$ if one of the two photons
is lost. The probability to produce a single detection in
mode $\tilde{a}$ for this term is approximately
$\beta^4\eta_t \eta_d$, since for long distances
$\eta_t \ll 1$ \cite{Explanation}. This detection creates the vacuum state
$|0\rangle$ for the remaining modes $a$ and $b.$ The state
created by the detection of a single photon in mode
$\tilde{a}$ is thus given by
\begin{equation}
\label{prepared_state}
\beta^2|0\rangle\langle0| + \alpha^2 |\psi\rangle\langle\psi|
\end{equation}
where
$|\psi\rangle=\frac{1}{\sqrt{2}}(a^{\dag}+b^{\dag})|0\rangle.$
The state $|\psi\rangle$ corresponds to an entangled state of the
two memories located at $A$ and $B$,
$\frac{1}{\sqrt{2}}\left(|1_{\rm A}0_{\rm B}\rangle+|0_{\rm
A}1_{\rm B}\rangle\right)$, while the vacuum state
$|0\rangle$ corresponds to $|0_A 0_B\rangle$, i.e. both
memories are empty. We emphasize that none of the three terms in Eq. (\ref{2photons}) leads to a component of the form $|1_A 1_B\rangle.$ This is a crucial difference compared to the pair-source protocols (DLCZ and P$^2$M$^3$).

The further steps are as for the DLCZ protocol
\cite{Duan01,Simon07}. Neighboring links are connected
via entanglement swapping, creating an entangled state
$\frac{1}{\sqrt{2}}\left(|1_{\rm A}0_{\rm Z}\rangle+|0_{\rm
A}1_{\rm Z}\rangle\right)$ between two distant locations
$A$ and $Z$. Moreover each location contains two memories,
denoted $A_1$ and $A_2$ for location $A$ etc. Entangled
states of the given type are established between $A_1$ and
$Z_1$, and between $A_2$ and $Z_2$. By post-selecting the
case where there is one excitation in each location, one
generates an effective state of the form
\begin{equation}
\label{postselected_state}
\frac{1}{\sqrt{2}}\left(|1_{\rm A1}1_{\rm Z2}\rangle+|1_{\rm A2}1_{\rm Z1}\rangle\right).
\end{equation}
The vacuum component in Eq. (\ref{prepared_state}) does not
contribute to this final state, since if one of the two pairs of
memories contains no excitation, it is impossible to detect
one excitation in each location. The vacuum components
thus have no impact on the fidelity of the final state. 
This is not the case for components involving two full memories as in Refs. \cite{Duan01, Simon07}, which may induce one excitation in each location and thus decrease the fidelity. 
Note that vacuum components, which exist for the single-photon source protocol already at the level of the elementary links,
occur for the pair-source protocols as well,
starting after the first entanglement swapping procedure \cite{Duan01}.

In contrast to the P$^2$M$^3$ protocol, the wavelength of the photons stored
in the memory is necessarily the same as that of the traveling photons,
requiring memories that operate in the wavelength range
where losses in optical fibers are minimal. Such memories
could be realized e.g. in Erbium-doped crystals \cite{Staudt07}, based on an off-resonant Raman process \cite{Raman}, electromagnetically induced transparency \cite{EIT}, or controlled reversible
inhomogeneous broadening \cite{Nilsson05}. 

The pair-source protocols require a fixed phase relationship 
between the two pair sources, i.e. between the $a^{\dag}a'^{\dag}$
and $b^{\dag}b'^{\dag}$ terms in Eq. (\ref{p2m3}) \cite{Simon07}.
There is no equivalent requirement for the single-photon source 
protocol, since the phase between the $a'^{\dag}b^{\dag}$ and $a^{\dag}b'^{\dag}$
terms in Eq. (\ref{2photons}) depends only on the beam-splitter transformation,
and not on the phase of the pump laser. It is important for all considered protocols that the photons from the two sources are indistinguishable, and that
the fiber lengths are stable on the time-scale of the entanglement creation for an elementary link \cite{Simon07}.\\

As we have indicated before, the absence of fundamental
errors proportional to the entanglement creation
probability leads to very significantly improved
entanglement distribution rates for the single-photon source protocol. We 
now discuss this improvement quantitatively. The time
required for a successful creation of an entangled state of
the form (\ref{postselected_state}) is given by
\cite{Simon07}
\begin{equation}
\label{total_time}
T_{\rm tot}=\left(\frac{3}{2}\right)^{n+1}\frac{L_0}{c} \frac{1}{P_0P_1...P_n P_{\rm pr}}.
\end{equation}
where $L_0=L/2^n$ is the length of an elementary link, $L$ is total distance and $n$ is the
nesting level of the repeater; $P_0$ is
the success probability for entanglement creation in an
elementary link; $P_i$ (with $i \geq 1$) is the success
probability for entanglement swapping at the $i$-th level,
and $P_{\rm pr}$ is the probability for a successful
projection onto the state Eq. (\ref{postselected_state}).

For the DLCZ protocol, the success
probability for entanglement creation in an elementary link
is $P_0=p\eta_t \eta_d$, while for the new single-photon
source protocol $P_0=2p_1\beta^2\eta_t \eta_d$. Here
$p_1$ is the probability that the source emits one photon ($p_1=1$ in the ideal case).
The weight of the vacuum component at each nesting level is larger in the
single-photon source protocol, and thus the success probabilities $P_i$
(with $i \geq 1)$ for entanglement swapping are somewhat
lower. However, the probability $P_0$ can be made much larger than
in the photon-pair source protocols. Overall, this leads to higher
entanglement distribution rates, as we detail now.

One can show from Eq. (\ref{total_time}) that the total
time required for entanglement distribution with the
single-photon protocol is \cite{Simon07bis}
\begin{equation}
\label{evaluation_time} T_{\rm
tot}=\frac{3^{n+1}}{2}\frac{L_0}{c}
\frac{\prod_{k=1}^{n}\left(2^k-\left(2^k-1\right)p_1\alpha^2\eta\right)}{\eta_d\eta_{t}p_1^{n+3}\beta^2\alpha^{2n+4} \eta^{n+2}}.
\end{equation}
Here $\eta=\eta_m\eta_d$, where $\eta_m$ is the memory
efficiency, and we assume photon-number resolving detectors
with efficiency $\eta_d$; $\eta_{t}=\exp{(-L_0/(2L_{\rm att}))}$ is the fiber
transmission efficiency, and $c=2\times10^8$ m/s is the
photon velocity in the fiber. To evaluate the potential
performance of our scheme, we calculate the average total
time for an entangled state distribution for a distance
$L=1000$ km with $L_{\rm att}=22$ km, corresponding to photons at the
telecommunications wavelength of 1.5 $\mu$m. We assume
$\eta_m=\eta_d=0.9$. High-efficiency memories
\cite{Highefficiencymemories} and photon-number-resolving detectors \cite{Kim99, Rosenberg05, Rosfjord} are
currently being developed. From Eq. (\ref{evaluation_time})
one can show that the optimal nesting level for our
protocol for these parameter values is $n=3,$ corresponding to $8$
elementary links. Assuming $p_1=0.95$ and optimizing the
formula (\ref{evaluation_time}) over $\beta$ gives $T_{\rm
tot}=250$ s with $\beta^2=0.11.$ For the DLCZ protocol, $T_{\rm tot}$ is also minimal for
$8$ links. The calculation of the errors due to
double-pair emission shows that in order to achieve a final
fidelity $F=0.9$ one has to choose $p=0.003$, leading to
$T_{\rm tot}=4600$ s. The new single-photon source protocol
thus reduces the average time for a successful distribution
of an entangled state over $1000$ km by a factor of 18.
This gain can be understood by comparing the values of
$P_0$ for the two protocols. For the new protocol, $P_0=0.01$, whereas for the DLCZ protocol
$P_0= 0.0001.$ The difference between the calculated factor 18 
and the ratio between the $P_0$ values is due to the lower success probabilities for entanglement swapping.\\

\begin{table}[htp]
\begin{ruledtabular}
\begin{tabular}{cccccc| |ccccc | cc cc cc | cc}
\multicolumn{8}{c} {Distance} & DLCZ (s)  & & $n$ & & SPS (s) & & $n$ & & $\beta^2$ & & gain \\ \hline
 & & & & 1000 km & & & & 4600 & & 3 & & 250 & & 3 & &  0.11 & &18 \\ 
 & & & &  1500 km & & & & 28400 & & 3 & & 1560  & & 3 & & 0.11 & &18 \\ 
 & & & & 2000 km & & & & 156700 & & 3 & & 6000  & & 4 & & 0.08 & &26  \\ 
 & & & & 2500 km & & & & 650000 & & 4 & & 15300 & & 4 & & 0.08 & &42 \\ 
\end{tabular} 
\end{ruledtabular}
\caption{Average times for entanglement distribution over various distances for the DLCZ and single-photon source (SPS) protocols. The optimal nesting level $n$ and beam-splitter transmission coefficient $\beta^2$ are given. We assume
high-efficiency memories and photon-detectors ($\eta_m=\eta_d=0.9$). For the DLCZ protocol, the fidelity of the final state constrains the probability $p$ of photon-pair emission to be small, e.g. for F=0.9, one has to choose $p=0.003$ for both $1000$ and $1500$ km and $p=7\times10^{-4}$ for both $2000$ and $2500$ km. In the case of single-photon source protocol, the fidelity is not fundamentally limited by the success probability of single-photon emission $p_1,$ chosen to be equal to $0.95.$ As shown in the last column, the gain for the single-photon source protocol compared to the DLCZ protocol increases from a factor of $18$ for $1000$ km to a factor of $42$ for $2500$ km. Note that the indicated average times could further be decreased by several orders of magnitude using e.g. multi-mode memories \cite{Simon07}. \label{table1}}
\end{table} 
Different distances from $1000$ to $2500$ km are considered and the corresponding  
gain with respect to the DLCZ protocol is shown in Table \ref{table1}. This gain increases with the distance. For example, it reaches a factor larger than 40 for 2500 km. The proposed protocol thus improves the entanglement distribution over long distances very significantly.

In our examples, we have chosen $p_1=0.95.$ The single-photon source protocol achieves on advantage over the DLCZ protocol as soon as $p_1>0.67,$ for all the considered distances between $1000$ km and $2500$ km. Efficient sources are thus required for profiting from the proposed protocol, cf. below.

Our architecture is compatible with
the use of multi-mode memories. The indicated average times
for entanglement creation can thus be reduced by several
orders of magnitude depending on the number of modes the
memory can store \cite{Simon07}. Spatial and frequency
multiplexing \cite{Collins07} and entanglement 
swapping by two-photon detection \cite{Jiang07} could further increase the
distribution rates.\\

We have shown that the single-photon protocol has no
fundamental error mechanism, i.e. the fidelity of the
created entangled states will be equal to one as long as
all components of the architecture work perfectly. However,
imperfections do affect the fidelity. We have studied two
kinds of imperfection that are likely to be relevant for
implementations, namely detector dark counts and a small
probability for a single-photon source to emit two
photons.

We first discuss dark counts, i.e. detector clicks in the
absence of photons. A dark count of one of the detectors
located at the central station can be associated with two
full memories, if the photons emitted by the two
sources located at A and B are in the modes $a$ and $b$.
The corresponding state $|1_A1_B\rangle$ does not coincide
with the expected entangled state and thus decreases the
fidelity. One can show by explicit calculation that the
fidelity of the final state compared to the ideal state
(\ref{postselected_state}) for a repeater with 8 elementary
links is \cite{Simon07bis}
\begin{equation}
\label{fidelity_darkcount}
F=1-16\Big[\frac{25}{\beta^2p_1}-\left(25\eta-1\right)\left(\frac{1}{p_1}-1\right)\Big]\frac{p_{\rm dark}}{\eta_t \eta_d}
\end{equation}
to first order in $\frac{p_{\rm dark}}{\eta_t \eta_d}$,
where $p_{\rm dark}$ is the probability for a detector to
give a dark count. Note that one only has to take into
account the effects of dark counts during the creation of
entanglement for the elementary links, where fiber
transmission losses are very large and real detection
probabilities correspondingly low. For the entanglement
swapping steps of the protocol the detectors are located
close to the memories and the real detection probabilities
are much larger than realistic dark count probabilities. We
evaluate the constraints on the dark counts by considering
the set of parameter values used before for a distance of
1000 km. One finds that in order to achieve $F=0.9$,
$p_{\rm dark}$ has to be smaller than $4.6\times 10^{-6}$.
This seems realistic. Transition-edge sensor detectors
can already resolve telecom-wavelength photons of $4$ ns duration at a repetition rate of $50$ KHz, with an efficiency of 0.88 and negligible noise 
\cite{Rosenberg05}. In the long run, NbN detectors promise to 
resolve even shorter pulses at higher rates \cite{Rosfjord}.

We now evaluate the tolerance of our scheme with respect to
undesired two-photon emissions by the sources. Two-photon emissions
cause errors that are similar in nature to those for dark counts. 
Such emissions might be due e.g. to pump laser scattering. One
shows by explicit calculation that for a repeater with 8
links the fidelity is given by \cite{Simon07bis}
\begin{equation}
\label{fidelity_2photon}
F=1-2\left(\frac{376}{p_1}-(1-\beta^2)(395\eta-19)\right)\frac{p_2}{p_1}
\end{equation}
to first order in $p_2$, where $p_2$ is the probability for
each source to emit two photons. For the same values as
above one finds that $p_2$ has to be smaller than
$3.7 \times 10^{-4}$ in order to achieve a fidelity
$F=0.9$. 

Single-photon sources as required for the presented
protocol, i.e. with high probability of single-photon emission and low
probability of two-photon emission can be realized with a variety of approaches. For
first demonstration experiments, the most promising
approach may be the use of asynchronous heralded
single-photon sources based on parametric down-conversion,
where $p_1>0.6$ and $p_2$ of order $10^{-4}$ have already
been achieved at $1.5$ $\mu$m \cite{Fasel04}. Even higher $p_1>0.8$
has been reported in Ref. \cite{Pittman05} at $780$ nm. In the long run, sources
based on quantum dots \cite{dots} embedded in microcavities
\cite{microcav} are likely to offer higher repetition
rates, which is important in order to fully profit from multi-mode memories. Quantum dot sources that emit at telecom wavelengths
are being developed \cite{Zinoni06}. Single atoms inside high-finesse cavities \cite{atoms_cavity} are also potential candidates, possibly combined with 
wavelength-conversion techniques \cite{Tanzilli05} in order to reach telecom wavelengths.

We have proposed a quantum repeater protocol based on
single-photon sources that eliminates the fundamental
errors due to double-pair emission which limit the
performance of previous protocols \cite{Duan01,Simon07}.
It is interesting to note that an important initial motivation for 
the development of single-photon sources was their application 
for point-to-point quantum key distribution (QKD), while quantum repeaters
were thought to require photon-pair sources. In fact, high-bit-rate point-to-point
QKD is achieved more conveniently using weak laser pulses with decoy state 
protocols \cite{decoy}. On the other hand, we have shown that single-photon sources
are very promising for the implementation of efficient quantum repeaters.

We acknowledge M. Afzelius for useful discussions and support from the EU integrated project Qubit applications (QAP) and the Swiss NCCR Quantum Photonics.

\end{document}